\documentclass{elsart3}

\usepackage{graphicx}
\usepackage{amssymb}

\def\NIMA#1#2#3{Nucl. Instr. Meth. Phys. Res. A {\bf #1}\ (#2)\ #3}

\def\IEEE#1#2#3{IEEE Trans. Nucl. Sci. vol. {\bf #1}\ (#2)\ #3}

\begin{document}

\begin{frontmatter}

\title{Electron identification performance with ALICE TRD prototypes}

\author[gsi]{A.~Andronic\thanksref{leave}},

\address[gsi]{Gesellschaft f{\"u}r Schwerionenforschung, Darmstadt, Germany;
Email:~A.Andronic@gsi.de}

{for the ALICE TRD Collaboration$^2$}

\thanks[leave]{On leave from NIPNE Bucharest, Romania.}
\thanks[info]{For the members of ALICE TRD Collab., see ref.~\cite{aa:obusch}}

\begin{abstract}
We present the electron/pion identification performance measured with
prototypes for ALICE TRD. 
Measured spectra of energy deposit of pions and electrons as well as 
their average values are presented and compared to calculations.
Various radiators are investigated over the momentum range of 1 to 6 GeV/c.
The time signature of TR is exploited in a bidimensional likelihood mothod.

\end{abstract}


\end{frontmatter}

\section{Introduction} \label{d:intro}

The ALICE Transition Radiation Detector (TRD) \cite{aa:tdr} is designed
for the study of quarkonia and open charm mesons in heavy-ion collisions 
at the CERN LHC.
The TRD will provide electron identification and particle tracking, 
in particular at the trigger level. 
A factor of 100 pion rejection for 90\% electron efficiency is the design 
goal of the detector and was demonstrated with earlier prototypes 
\cite{aa:i3e,aa:tdr}.  


\vspace{-3mm}
\section{Experimental setup} \label{d:meth} 

The measurements are performed with four identical prototype drift 
chambers (DC), with a construction similar to that of the final ALICE 
TRD \cite{aa:tdr}, but of a smaller active area (25$\times$32~cm$^2$). 
The drift chambers have a drift region of 30~mm and an amplification region 
of 7~mm.
The entrance window (25~$\mu$m aluminized Kapton) simultaneously serves  
as gas barrier and as drift electrode.
For the final detectors this window will be glued on the radiator wall, 
but for the prototypes the window is part of the drift chamber to allow 
tests of various radiators.
We operate the drift chambers with the standard gas mixture for the TRD, 
Xe,CO$_2$(15\%), at atmospheric pressure.
The gas is recirculated using a dedicated gas system.
The gas gain is about 4000.

Our radiators are sandwiches composed of polypropylene (PP) fibre mats 
(of 5~mm thickness each) and two sheets of Rohacell foam of thickness 6~mm 
(INV6, AIK6, which contain 8 fibre mats) or 8~mm (INV8, which contains 
only 7 mats). 
A carbon fibre reinforcement  of about 100~$\mu$m thickness is applied on 
the external sides of the Rohacell sheets. INV and AIK stand for two 
different manufacturers of this coating.
The total thickness of such a radiator is 4.8~cm. In addition, 
for reference purposes, we tested pure PP fibres radiator of 4~cm thickness.

We use a prototype of the charge-sensitive preamplifier/shaper (PASA) 
especially designed and built for the TRD in 0.35~$\mu$m CMOS technology. 
It has a noise on detector of about 1000 electrons r.m.s. and the FWHM 
of the output pulse is about 100~ns for an input step function.
The nominal gain of the PASA is 12~mV/fC, but during the present measurements 
we use a gain of 6~mV/fC for a better match to the range of the employed 
Flash ADC (FADC) system with 0.6~V voltage swing.
The FADC has adjustable baseline, an 8-bit non-linear conversion
and 20~MHz sampling frequency.
The FADC sampling was rebinned in the off-line analysis to obtain 100~ns 
time bins as for the final ALICE TRD \cite{aa:tdr}. 

The measurements are carried out with four identical layers radiator/detector
at beam momenta of 1, 1.5, 2, 3, 4, 5, and 6~GeV/c at the T10 secondary 
beamline of the CERN PS. 
The momentum resolution is $\Delta p/p\simeq 1\%$.
The beam is a mixture of electrons and negative pions.
Similar sample sizes of pion and electron events are acquired within the
same run via dedicated triggers.
We select clean samples of pions and electrons using coincident thresholds 
on two Cherenkov detectors and on a lead-glass calorimeter \cite{aa:i3e}.
The incident angle of the beam with respect to the drift direction is 
15$^\circ$ to avoid gas gain saturation due to space charge \cite{aa:gain}.

\vspace{-5mm}
\section{General detector performance}

\begin{figure}[htb]
\vspace{-.8cm}
\centering\includegraphics[width=.5\textwidth,height=.45\textwidth]{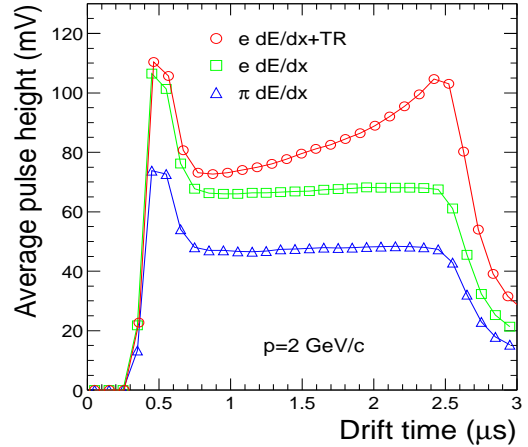}
\vspace{-.8cm}
\caption{Average pulse height as a function of drift time for pions 
and electrons (with and without radiator).}
\label{aa:f1}
\end{figure}

In Fig.~\ref{aa:f1} we present the measured average signals as a function 
of drift time for pions and electrons (with and without radiator),
for the momentum of 2~GeV/c. For our nominal drift field of 0.7 kV/cm, 
the detector signal is spread over about 2~$\mu$s (the time zero is 
arbitrarily shifted).
The peak at small drift times originates form the amplification region,
while the plateau is from the drift region.
For the electrons, when using a radiator, the contribution of TR, which is 
preferentially absorbed at the entrance of the detector, is evident.

\begin{figure}[htb]
\vspace{-.7cm}
\centering\includegraphics[width=.53\textwidth,height=.49\textwidth]{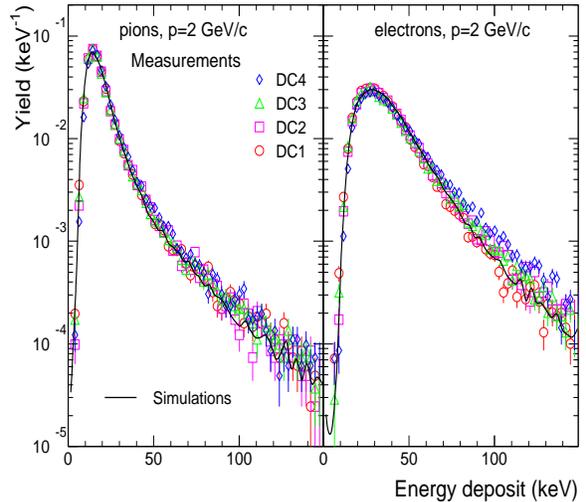}
\vspace{-.7cm}
\caption{Spectra of the energy deposit of pions and electrons (INV6 radiator) 
for the momentum of 2~GeV/c. The symbols represent the measurements, the lines 
are calculations.}
\label{aa:f2}
\end{figure}

In Fig.~\ref{aa:f2} we present the measured distributions of integrated 
energy deposit for pions and electrons for the momentum of 2~GeV/c, 
for each of the four layers. 
The data are compared to calculations, which include ionization 
energy loss \cite{aa:dedx} and, in case of electrons, transition radiation 
(TR). 
For TR production we employ a parametrization of a regular radiator 
\cite{aa:fab}, which we tune to describe the electron spectra.
The calculations include TR absorption in the radiator reinforcement
as well as in the detector volume.
As seen in Fig.~\ref{aa:f2}, we can achieve a good agreement with 
the measurements with a reasonable (but not unique) set of parameters 
("foil" thickness $d_1$=10~$\mu$m, gap $d_2$=80~$\mu$m, number of "foils" 
$N_f$=270). Pure TR measuremens are described equally well \cite{aa:obusch}.

\begin{figure}[htb]
\vspace{-.9cm}
\centering\includegraphics[width=.53\textwidth,height=.49\textwidth]{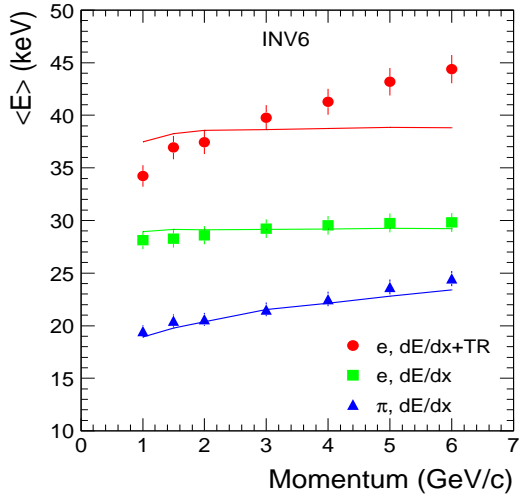}
\vspace{-.8cm}
\caption{Average integrated energy of pions and electrons as a function of 
momentum. The symbols represent the measurements, the lines are calculations.}
\label{aa:f3}
\end{figure}

However, with this recipe we cannot consistently describe our measurements 
over all the momentum range, as seen in Fig.~\ref{aa:f3}, where we compare
the measured average integrated energy with simulations. 
The experimental errors represent an estimated 3\% value. 
In addition, we estimate an overall 5\% systematic uncertainty of the energy 
calibration.
The agreement on dE/dx is very good, both for pions and for electrons
\cite{aa:dedx}.
The momentum dependence of the TR yield is not properly described. 
TR is overestimated at low momenta and underestimated at high momenta. 
Other sets of $d_1$/$d_2$/$N_f$ values also fail to reproduce the data.
An implementation of an irregular radiator is in progress.

\section{Electron/pion identification} \label{d:res}

\begin{figure}[hbt]
\vspace{-.8cm}
\centering\includegraphics[width=.52\textwidth,height=.49\textwidth]{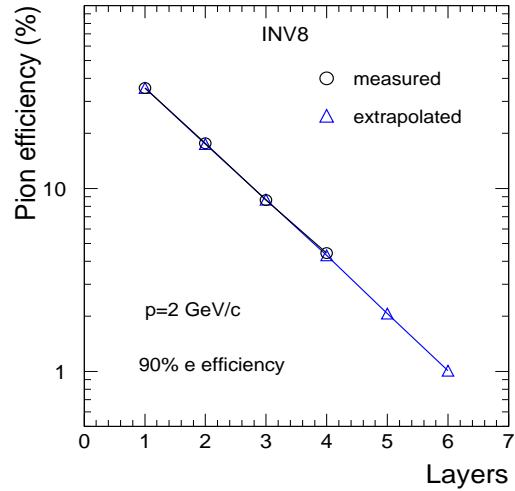}
\vspace{-.8cm}
\caption{Pion efficiency as a function of the number of layers included 
in the likelihood.} 
\label{aa:f4}
\end{figure}

In Fig.~\ref{aa:f4} we present the pion efficiency as a function of the
number of layers considered in the likelihood.
Measured data (open circles) are compared to the results of calculations 
using as inputs the measured single layer performance.
As expected, the agreement between the two cases is very good.
Consequently, for the following, the expected pion efficiency for 
six layers (final configuration in ALICE) is calculated from the 
single-layer variables for which the measured signals of the four 
layers are added to improve statistics.

\begin{figure}[hbt]
\vspace{-.9cm}
\centering\includegraphics[width=.54\textwidth,height=.7\textwidth]{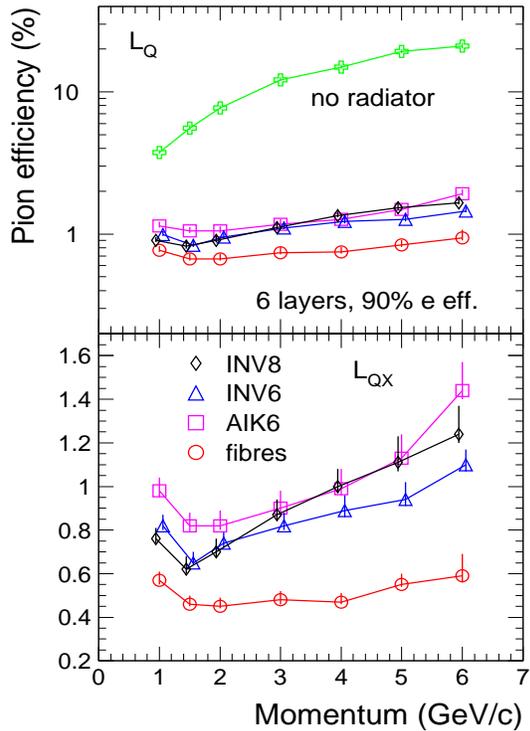}
\vspace{-.5cm}
\caption{Pion efficiency as a function of momentum for various radiator
types. Upper panel: likelihood on total charge; for this case the performance 
in case of dE/dx only (no radiators) is included. Lower panel: bidimensional 
likelihood on charge and position (see text).} 
\label{aa:f5}
\end{figure}

In Fig.~\ref{aa:f5} we present the pion efficiency (the pion rejection 
is the inverse of this value) as a function of momentum for three 
sandwich radiators and for pure PP fibres of 4~cm thickness.
The upper panel shows the results for a likelihood method on integrated 
charge (L$_Q$). For this case the results without radiator are included 
as well, showing a strong degradation as a function of momentum due to 
the pion relativistic rise. Transition radiation nicely compensates this, 
leading to a pion rejection factor of around 100 at 90\% electron efficiency,
with a weak dependence on momentum. 
The three coated sandwich radiators show a similar performance, while 
the pure fibres radiator is better, although it is thinner by 0.8~cm.

The lower panel of Fig.~\ref{aa:f5} is for a so-called bidimensional 
likelihood \cite{aa:hol}, for which the distribution of the time bin with 
the maximum measured amplitude is used together with the integrated charge 
measurement (L$_{QX}$). 
An improvement of the pion rejection by a factor of about 1.6 is achieved 
with this method.
This improvement is needed to provide a safety factor for the expected 
degradation of the performance in the high-multiplicity heavy-ion collisions 
at the LHC \cite{aa:tdr}.
The pion rejection can be further enhanced by exploiting the amplitude 
measurement in each time bin, which is, however, highly correlated
and cannot be used in a straightforward way in a likelihood method.
Developments in this direction are under way.

\begin{figure}[hbt]
\vspace{-.7cm}
\centering\includegraphics[width=.52\textwidth,height=.46\textwidth]{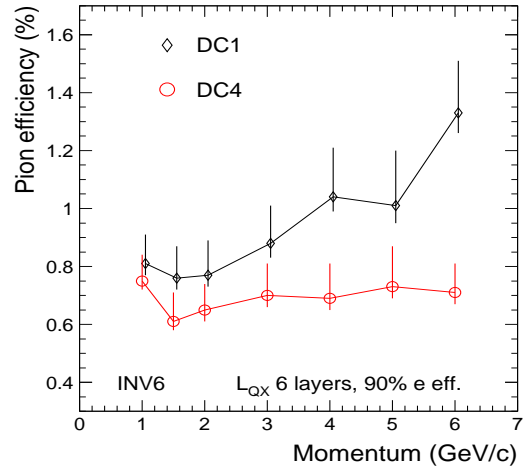}
\vspace{-.8cm}
\caption{Pion efficiency as a function of momentum for layer 1 and layer 4
using the bidimensional likelihood method.} 
\label{aa:f6}
\end{figure}

The pion efficiencies presented above are measured averages over 4 layers.
We have observed that the pion rejection performance extracted from 
individual layers improves substantially from layer 1 to layer 4. 
This is also apparent in Fig~\ref{aa:f2} in the charge spectra of electrons.
As seen in Fig.~\ref{aa:f6}, this improvement is bigger for larger momenta. 
Propagation and absorption of TR deeper into the stack of detectors
could explain this behavior.

\section{Summary} \label{aa:sum}

We have demonstrated in measurements with prototypes that the
requested pion rejection factor of 100 can be achieved for ALICE TRD.
The measurements of energy deposit of pions and electrons were compared 
to calculations.
Various radiators have been investigated over the momentum range 1-6~GeV/c.
A bidimensional likelihood method sizeably enhances the pion rejection and 
is under current study for further improvements.


\end{document}